\documentclass{elsart}

\usepackage{graphicx}
\usepackage{ifpdf}
\usepackage{graphicx,amssymb,lineno}
\ifpdf
\usepackage[%
  pdftitle={Instructions for use of the document class
    elsart},%
  pdfauthor={Simon Pepping},%
  pdfsubject={The preprint document class elsart},%
  pdfkeywords={instructions for use, elsart, document class},%
  pdfstartview=FitH,%
  bookmarks=true,%
  bookmarksopen=true,%
  breaklinks=true,%
  colorlinks=true,%
  linkcolor=blue,anchorcolor=blue,%
  citecolor=blue,filecolor=blue,%
  menucolor=blue,pagecolor=blue,%
  urlcolor=blue]{hyperref}
\else
\usepackage[%
  breaklinks=true,%
  colorlinks=true,%
  linkcolor=blue,anchorcolor=blue,%
  citecolor=blue,filecolor=blue,%
  menucolor=blue,pagecolor=blue,%
  urlcolor=blue]{hyperref}
\fi

\makeatletter
\def\elsartstyle{%
    \def\normalsize{\@setfontsize\normalsize\@xiipt{14.5}}
    \def\small{\@setfontsize\small\@xipt{13.6}}
    \let\footnotesize=\small
    \def\large{\@setfontsize\large\@xivpt{18}}
    \def\Large{\@setfontsize\Large\@xviipt{22}}
    \skip\@mpfootins = 18\p@ \@plus 2\p@
    \normalsize
}
\@ifundefined{square}{}{}
\makeatother

\usepackage{amssymb}
\usepackage{graphicx}
\usepackage{dcolumn}
\usepackage{bm}
\usepackage{longtable}

\begin{document}
\begin{frontmatter}

\title{Experimental and First principle calculation of Co$_x$Ni$_{(1-x)}$Si solid solution structural stability}

\author[1]{J. Teyssier},
\ead{Jeremie.Teyssier@physics.unige.ch\\phone: +41 22379 60 76\\fax: +41 22379 60 76}
\author[1]{R. Viennois},
\author[1]{J. Salamin},
\author[1]{E. Giannini} and
\author[1]{D. van der Marel}

\address[1]{D\'epartement de Physique de la Mati\`ere Condens\'ee, Universit\'e de Gen\`eve, Quai Ernest-Ansermet 24, 1211 Gen\`eve 4, Switzerland}

\date{\today}

\begin{abstract}

We report the investigation of the structural stability of Co$_{(1-x)}$Ni$_x$Si monosilicides for $0<x<1$. As CoSi crystallizes in the FeSi-type structure (B20) and NiSi is stable in the MnP-type structure (B31), a complete set of samples has been synthesized and a systematic study of phase formation under different annealing conditions were carried out in order to understand the reason of such a structural transition when x goes from 0 to 1. This study has revealed a limit in the solubility of Ni in CoSi B20 structure of about 17.5 at.\% and of Co in NiSi B31 phase of about 13 at.\%. For $0.35<x<0.74$ both B20 and B31 phases are present in the sample at there respective limits of solubility. The temperature dependence of the magnetic susceptibility has also been measured revealing diamagnetic behaviors. Optimal structural parameters and phase stability of the solid solution have been investigated using self-consistent full-potential linearized augmented plane wave method (FP-LAPW) based on the density functional theory (DFT). This calculation well predicts the structural instability observed experimentally.

\end{abstract}
\begin{keyword}
A intermetallics \sep B crystal growth \sep C crystal structure \sep C phase transitions


\end{keyword}
\end{frontmatter}

\section{Introduction}

During the past years, the transition metal monosilicides MSi with B20 cubic structure have attracted lot of attentions due to their interesting and various ground states. Notably, MnSi is an itinerant helimagnetic metal for $T<30K$ \cite{grigoriev2006}, FeSi is a paramagnetic Kondo insulator \cite{Paschen1997}, CrSi is a Pauli paramagnetic metal \cite{Shinoda1966,Wernick1972} and CoSi is a diamagnetic metal \cite{Wernick1972,Han2001,Nakanishi1980}. Moreover, it should be noted that the solid solution Fe$_x$Co$_{(1-x)}$Si also exhibits itinerant helimagnetic metallic behavior like MnSi for $0.4<x<0.9$ ($T_c=60K$ for $x=0.6$) although the two end-compounds FeSi and CoSi have no magnetic ordering \cite{Moriya1973,Manyala2000,Manyala2004,Onose2005}. On the other hand, NiSi is a diamagnetic metal which crystallizes in the B31 orthorhombic structure at room pressure \cite{Wernick1972,Meyer1997}. Thus, it is interesting to study the solid solution Co$_{(1-x)}$Ni$_x$Si for which no physical properties has been reported yet. Although the CoSi-NiSi phase diagram has already been investigated, very different limits of the solubility of Ni in the B 20 CoSi phase from 10 to 50\% has been reported \cite{McNeill1964,Watanabe1979,Alekseeva1982} depending on synthesis conditions and thermal treatments. On the other side of the phase diagram, only the Co$_{0.2}$Ni$_{0.8}$Si in the B31 structure was synthesized\cite{Wittmann1961}.

The isothermal cross-section of the ternary phase diagram Co-Ni-Si at 800 $^oC$ was reported by van Beek et al. \cite{vanbeek2000}. Even though they investigated only the solid-solid equilibria at one given temperature they first reported the existence of a miscibility gap between CoSi and NiSi.

In this paper, we report theoretical and experimental study of the structural, thermodynamic and magnetic properties of the whole range of Co$_{(1-x)}$Ni$_x$Si compositions. The limits of solubility of Ni in B20 CoSi and Co in B31 NiSi are determined by combining X-ray diffraction, differential thermal analysis and quantitative EDX chemical analysis. We show that the stability ranges of the B31 and B20 structures are well predicted by total energy LDA calculations.

\section{Crystal structure}

CoSi crystallizes in the common FeSi-type B20 structure (P2$_1$3 (198))\cite{Boren1934}. NiSi crystallizes in the MnP B31 structure (Pnma (62))\cite{Rabadanov2002}. The existence of NiSi with the B20 structure has only been reported when part of the silicon is substituted with Al \cite{Esslinger57}.

\begin{figure}[!h]
  \center \includegraphics[height=3cm]{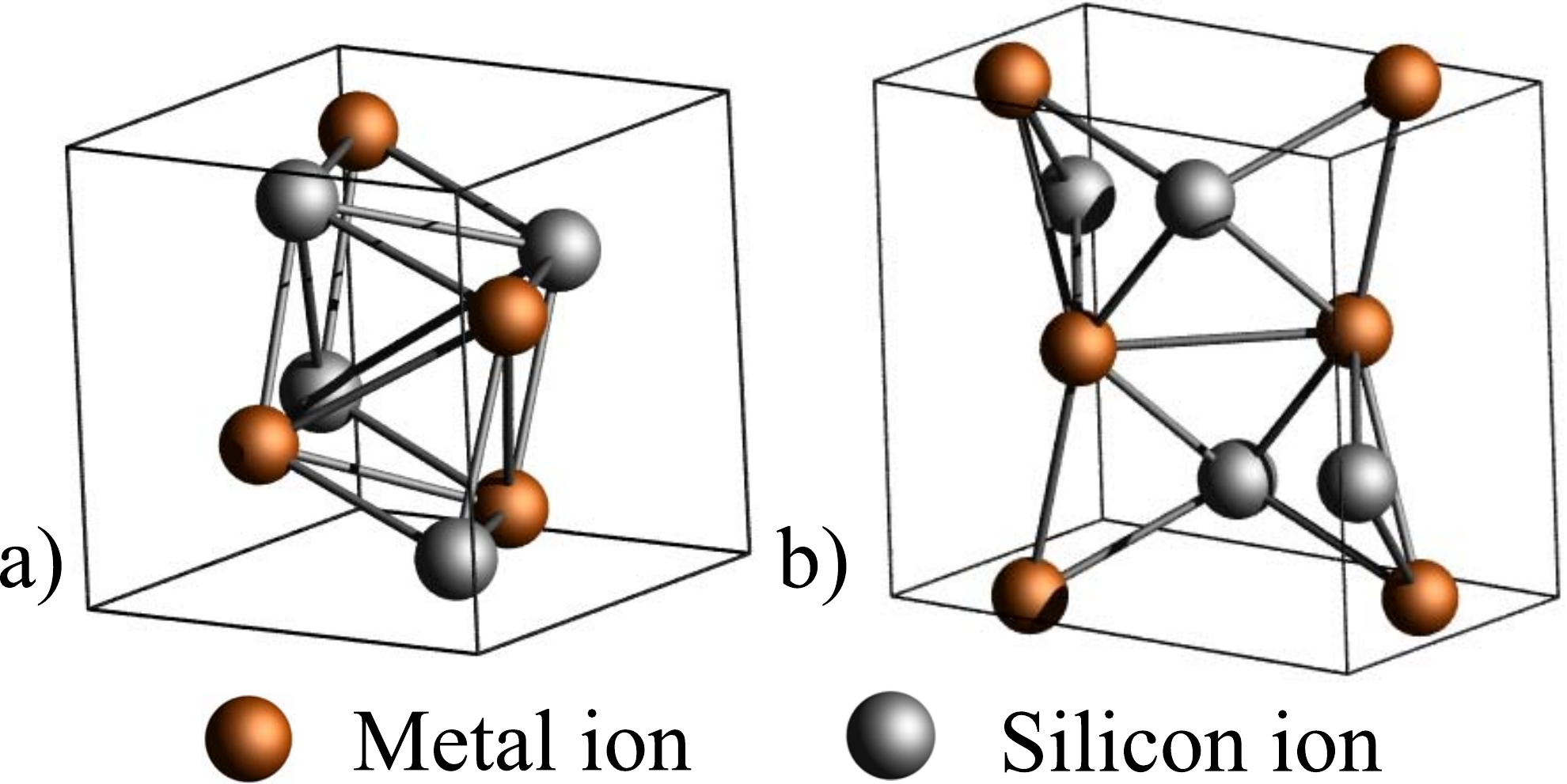}\\
  \caption{a) B20 structure of CoSi, b) B31 structure of NiSi}\label{structure}
\end{figure}

 The positions of both Metal and Si atoms in the B20 unit cell are ($x$,$x$,$x$), and the 3 permutations of ($x+\frac{1}{2}$,$\frac{1}{2}-x$,$\overline{x}$). In CoSi, these values are found to be $x_{Co}$=0.14 and $x_{Si}$=0.843 and the lattice parameter a=4.438 {\AA} \cite{Boren1934}. In the B31 structure, the positions of both Metal and Si atoms are ($x$,$\frac{1}{4}$,$z$), ($\overline{x}+\frac{1}{2}$,$\frac{3}{4}$,$z+\frac{1}{2}$), ($\overline{x}$,$\frac{3}{4}$,$\overline{z}$) and ($x+\frac{1}{2}$,$\frac{1}{4}$,$\overline{z}+\frac{1}{2}$). For NiSi, these values are found to be $x_{Co}$=0.00757, $z_{Co}$=0.18772,$x_{Si}$=0.32090 and $z_{Si}$=0.08168. Lattice parameters are a=5.194 {\AA}, b=3.323 {\AA} and c=5.629 {\AA} \cite{Rabadanov2002}.

\begin{figure}[!h]
   \center\includegraphics[width=8cm]{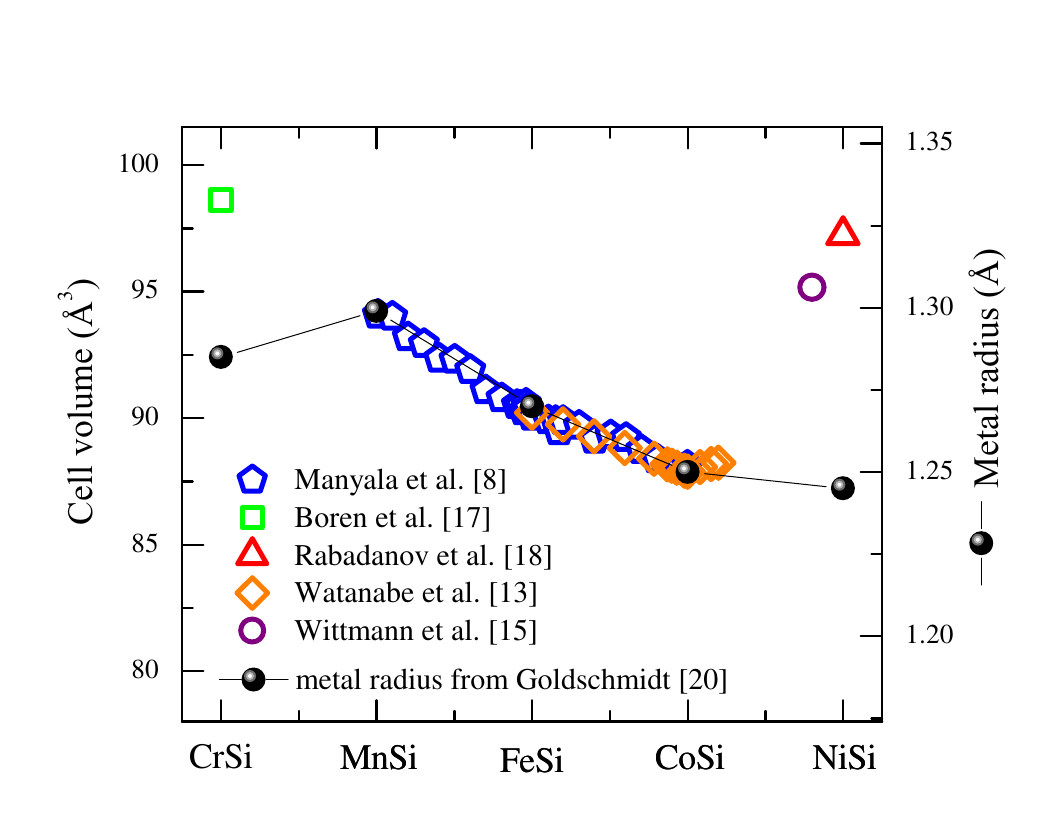}\\
  \caption{Evolution of the volume of the unit cell with chemical composition of monosilicides. Atomic radius values from Goldschmidt \cite{Goldschmidt1928} are given as an indication.}\label{cell_vol}
\end{figure}

In the solid solutions of transition metal (Mn$\rightarrow$Fe$\rightarrow$Co) monosilicides, the cell parameter decreases, following the evolution of the ionic radius of the transition metal ion (Fig.~\ref{cell_vol}).

A trend inversion, already reported by Watanabe et al. \cite{Watanabe1979} (orange diamonds in Fig.~\ref{cell_vol}), occurs between CoSi and NiSi for which the lattice parameter increases whereas the ionic radius of the metal ion is still decreasing. They noticed this evolution of the structure of Co$_{(1-x)}$Ni$_x$Si solid solution until $x\sim0.2$.\\

\section{Experimental}

All samples were synthesized using a home made arc furnace with a water cooled copper crucible, starting from 4N purity transition metals and 6N silicon chunks. An annealing at 900$^oC$ from a minimum of 12 hours to a maximum of a week under high vacuum (about $5.10^{-7}$ mbars) is necessary to improve the crystalline order, increase the limits of solubility and decrease the deviation on x in solid solutions. This annealing temperature was chosen close to the melting point of NiSi ($T=979^oC$ \cite{Richter2004}, $T=982^oC$ \cite{Du1999}). Samples were processed every $x=0.1$.

X-ray powder diffraction (XRD) was performed in a Philips PW1820 diffractometer using the $K_{\alpha}$ radiation of a Cu tube ($\lambda=1.5406$ {\AA}). The XRD spectra were analyzed with a full pattern profile refinement method using the Fullprof program suite\cite{Fullprof}.

The stability range and melting temperature of each composition where measured by Differential Thermal Analysis (DTA) in a SETARAM TAG 24 thermal analyzer using Al$_2$O$_3$ as a reference. DTA measurements were performed under flowing Ar (0.6 l/h). About 50 mg of each sample were subjected twice to the same run at 5 $^o$C/min. The melting point of end compounds was identified as the onset of the second heating endothermic peak. The Liquidus temperatures for solid solutions and mixed phased samples were assumed to be the offset of the second broad endothermic peak during the second heating run. The analysis of the chemical composition was carried out by Electron Dispersive X-ray spectroscopy (EDX) in a LEO 438VP electron microscope coupled to a Noran Pioneer X-ray detector, at a beam energy of 20 KeV. Quantification of elements was done on the K-lines using internal calibration.

\section{Calculation of structural stability}\label{details_calculation}

The \emph{ab initio} structural optimizations and total energy calculations were carried out using the density functional theory DFT method as implemented in the Quantum-ESPRESSO package \cite{QE}. The calculations were carried out using an exchange-correlation functional by Perdew, Burke, and Ernzerhof \cite{Perdew1996} and Vanderbilt-type ultrasoft pseudopotentials \cite{Vanderbilt1990}. Wave functions were expanded in a plane wave basis to a 80 Ry cutoff. A kinetic energy cutoff of 200 Ry was used for the charge density. A 3x3x3 k-mesh has been used to sample the Brillouin zone.
First, the atomic positions and lattice constants were optimized for CoSi, Co$_{0.5}$Ni$_{0.5}$Si and NiSi in both B20 and B31 crystal structure. The starting point was the parameters extracted from X-ray refinements when available. For CoSi and Co$_{0.5}$Ni$_{0.5}$Si in the B31 configuration we took the parameters of NiSi and for NiSi and Co$_{0.5}$Ni$_{0.5}$Si in the B20 configuration, we took parameters of CoSi.

In table \ref{comparaison}, we report a summary of the lattice parameters obtained experimentally in this work from powder diffraction, from structural optimization and from the literature.
\begin{center}
\begin{table}[p!]
  \begin{tabular}{|c|c|c|c|c|c|c|c|c|c|c|}
    \hline
    &\multicolumn{3}{c|}{B20}&\multicolumn{7}{c|}{B31}\\
    \hline
    Compound&a&$x_M$&$x_{Si}$&a&b&c&$x_M$&$z_M$&$x_{Si}$&$z_{Si}$\\
    \hline
    CoSi exp.&4.444&0.143&0.844&&&&&&&\\
    CoSi opt.&4.442&0.143&0.843&5.335&2.910&6.009&0.0047&0.1964&0.3126&0.0582\\
    CoSi \cite{Boren1934}&4.438&0.140&0.843&&&&&&&\\
    \hline
    NiSi exp.&&&&5.1818&3.334&5.619&0.0069&0.1876&0.3122&0.1041\\
    NiSi opt.&4.515&0.146&0.846&5.2972&3.2549&5.6997&0.0068&0.1884&0.3181&0.0786\\
    NiSi \cite{Rabadanov2002}&&&&5.194&3.323&5.629&0.0076&0.1877&0.3209&0.0817\\
    \hline
    Co$_{0.5}$Ni$_{0.5}$Si opt.&4.475&0.144&0.846&5.369&3.029&5.910&0.0053&0.1929&0.3096&0.0640\\
    \hline
  \end{tabular}

  \caption{Comparison of experimental and calculated lattice parameters and atomic positions for B20 and B31 structures for CoSi, Co$_{0.5}$Ni$_{0.5}$Si and NiSi. \emph{exp.} and \emph{opt.} refer respectively to experimental and ab-initio optimized structures from this study.}\label{comparaison}
\end{table}
\end{center}

The total energy of Co$_{(1-x)}$Ni$_x$Si was then calculated for $x=0, 0.25,0.5,0.75$ and $1$. The calculation has been performed using the experimental parameters when available. When the cell parameters were not experimentally accessible, we used the ones from structural optimization calculations. For the B20 structure, the matching between calculated and experimental parameters is good in the range where both are available ($0<x<0.35$) (green open circles in Fig.~\ref{exp_lda}b). This is not the case for the B31 structure where LDA overestimates the cell parameter (green closed circles in Fig.\ref{exp_lda}b). For the B31 structure, we used the composition dependence of the cell parameter obtained from calculation scaled to match the experimental values in the range $0.74<x<1$ (gray closed squares in Fig.~\ref{exp_lda}b).\\

 The self consistent total energy calculation has been performed on a 10x10x10 k-mesh of the irreducible Brillouin Zone. For $x=0.25$, $x=0.5$ and $x=0.75$, we used a unit cell where respectively $\frac{1}{4}$, $\frac{2}{4}$ and $\frac{3}{4}$ Co atoms where substituted by Ni in both B20 and B31 structures.

\begin{figure}[!h]
   \center\includegraphics[width=8cm]{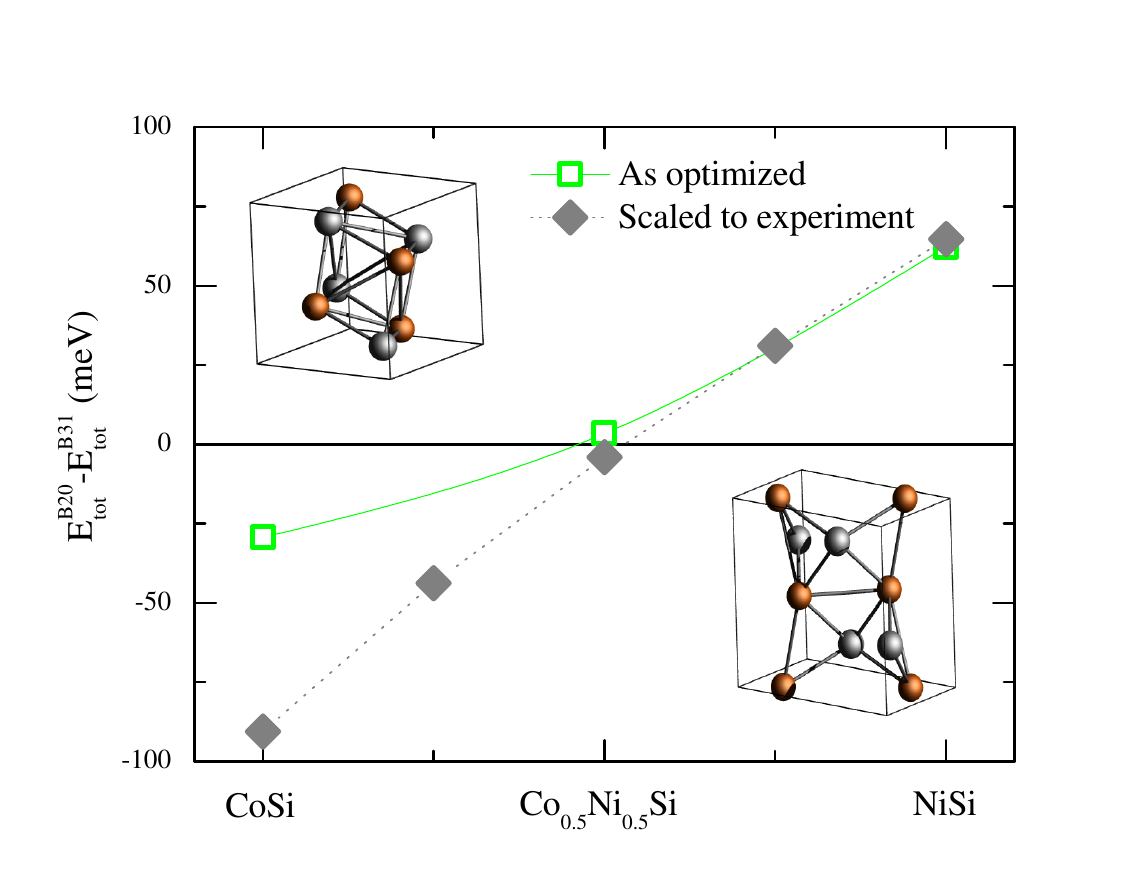}\\
  \caption{(color online) Open and closed symbols correspond to the total energy calculated from theoretical structural parameters and experimental refinements respectively.}\label{stab_lda}
\end{figure}

The realistic estimation of the value of the cell parameters allows us to calculate the total energy of all structures (real and hypothetic). In order to show how sensitive is the total energy to the deviation in cell parameters, we plotted in Fig.~\ref{stab_lda} the total energy difference on both "as-optimized" (open symbols) and "experimental" (closed symbols) structures.

 The total energy difference between B20 and B31 phases (Fig.~\ref{stab_lda}) shows that the B20 structure is stable up to a concentration $x\simeq0.55$. Above this concentration, the B31 configuration saves energy. Even if this calculation cannot predict the limit of solubility, the domain of stability of the B20 (Ni in CoSi) is larger than the one of B31 phase (Co in NiSi). This agrees well with the experimental observation of a much larger limit of miscibility of Ni in B20 CoSi than Co in B31 NiSi. In the following section, we discuss the experimental results.

\section{Experimental results}

From systematic XRD, SEM and EDX analyses of the whole series of samples, we have identified three different composition ranges, characterized by three different sample morphologies when $x$ is ranging from 0 to 1:
\begin{itemize}
  \item{a pure B20 solid solution;}
  \item{a mixture of both B20 and B31 solid solutions;}
  \item{a pure B31 solid solution.}
\end{itemize}

\begin{figure}[!h]
   \center\includegraphics[width=8cm]{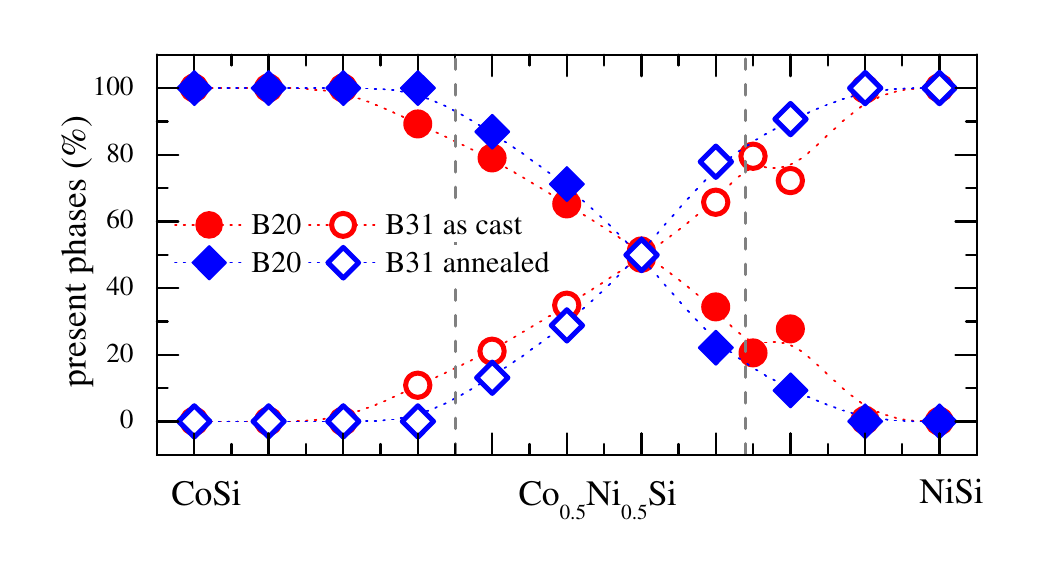}\\
  \caption{B20 and B31 phase molar fractions present in the samples before and after annealing. The abscissa is the nominal composition.}\label{sol}
\end{figure}

From XRD pattern refinement, the cell parameters and the ratio of the two different crystalline forms were extracted. The amount of the two different phases as a function of the nominal composition is plotted in Fig\ref{sol}. It clearly appears that below a certain level of substitution (Ni in B20-CoSi and Co in B31-NiSi) a good solubility is observed and samples are single phased.

\begin{figure}[!h]
   \center\includegraphics[width=8cm]{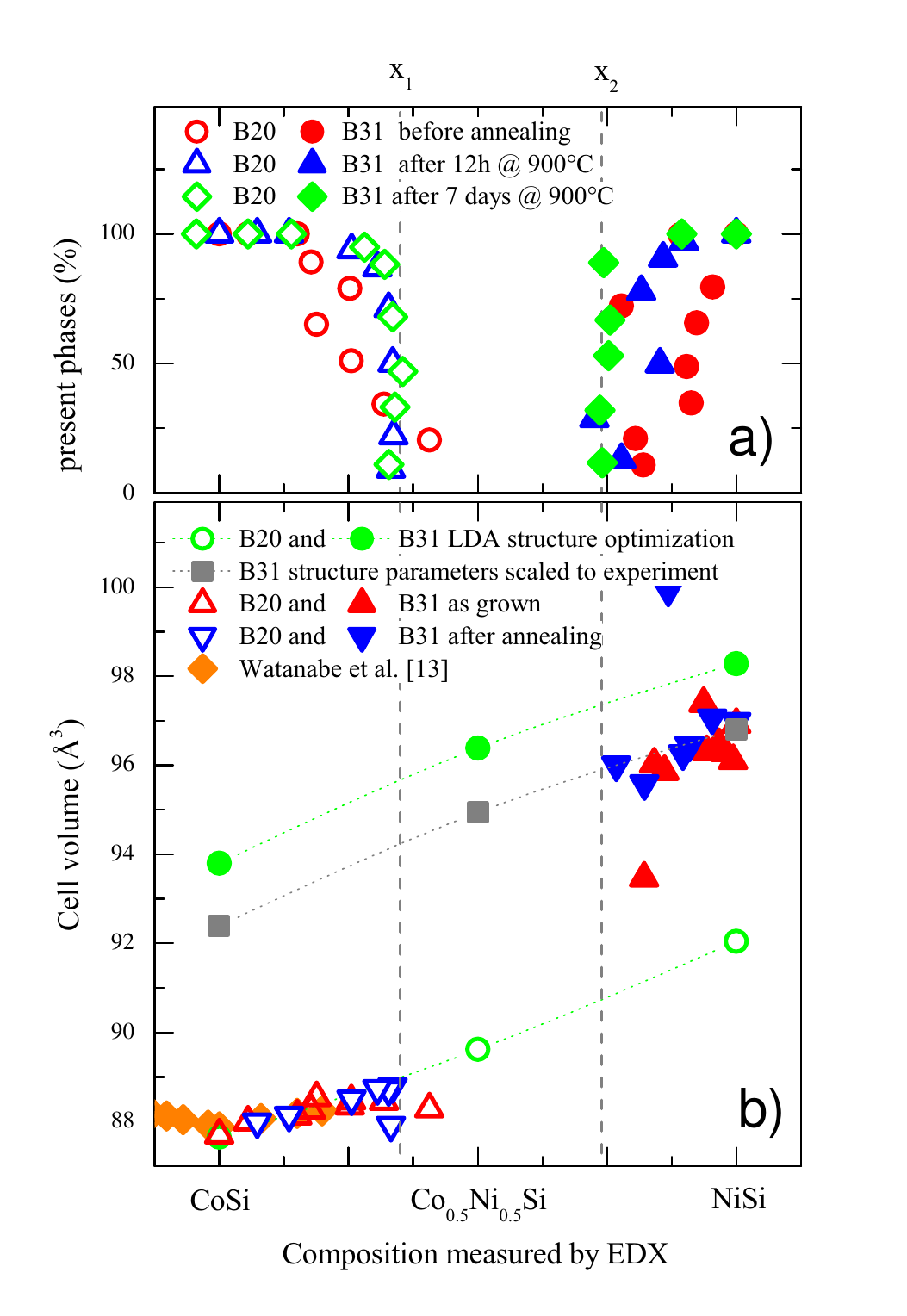}\\
  \caption{a) Molar fraction of the B20 and B31 phases in the sample before and after annealing. b)Experimental and LDA predicted values of the cell volume for B20 (closed symboles) and B31 (open symbols) structures. The vertical lines represent the two limites of solubility of Ni in CoSi and Co in NiSi.}\label{exp_lda}
\end{figure}

By combining the XRD with EDX analyses, it is possible to determine the exact composition and the cell parameters of each solid solution present in the multiphase mixture with $0.3<x<0.9$. The amount of each phase as a function of real and nominal compositions, is shown in Fig.~\ref{exp_lda}~a and Fig.~\ref{sol}, respectively. The evolution of the cell volume (i.e. average cell parameter) with the real composition determined from EDX is compared to LDA structural optimizations in Fig.~\ref{exp_lda}b.

In the limit of sensitivity of the XRD technique, the Co$_{(1-x)}$Ni$_x$Si B20 phase is observed for $x<x_1=0.35$ (17.5 at.\%) and the B31 phase is observed for $x>x_2=0.74$ (13 at.\%). These two limits of solubility are displayed as vertical dashed lines in Fig.~\ref{exp_lda} and \ref{sol}. We noticed a good agreement with van Beek et al. \cite{vanbeek2000} who reported limits of solubility at 800$^oC$ of 12 at.\% of Co in NiSi and 22 at.\% of Ni in CoSi. We also learn from Fig.~\ref{sol}, in which these limits of solubility almost coincide with the limit where the samples present a single phase, that the solid solutions are stable up to the limit of solubility.

Within the stability range, LDA structural predictions for the B20 structure type agree perfectly with the experimental parameters. For the B31 structure, energy minimum is found for a volume cell slightly higher than what is observed experimentally. A distorsion along the c axis is also observed.

\begin{figure}[h!]
   \center\includegraphics[width=10cm]{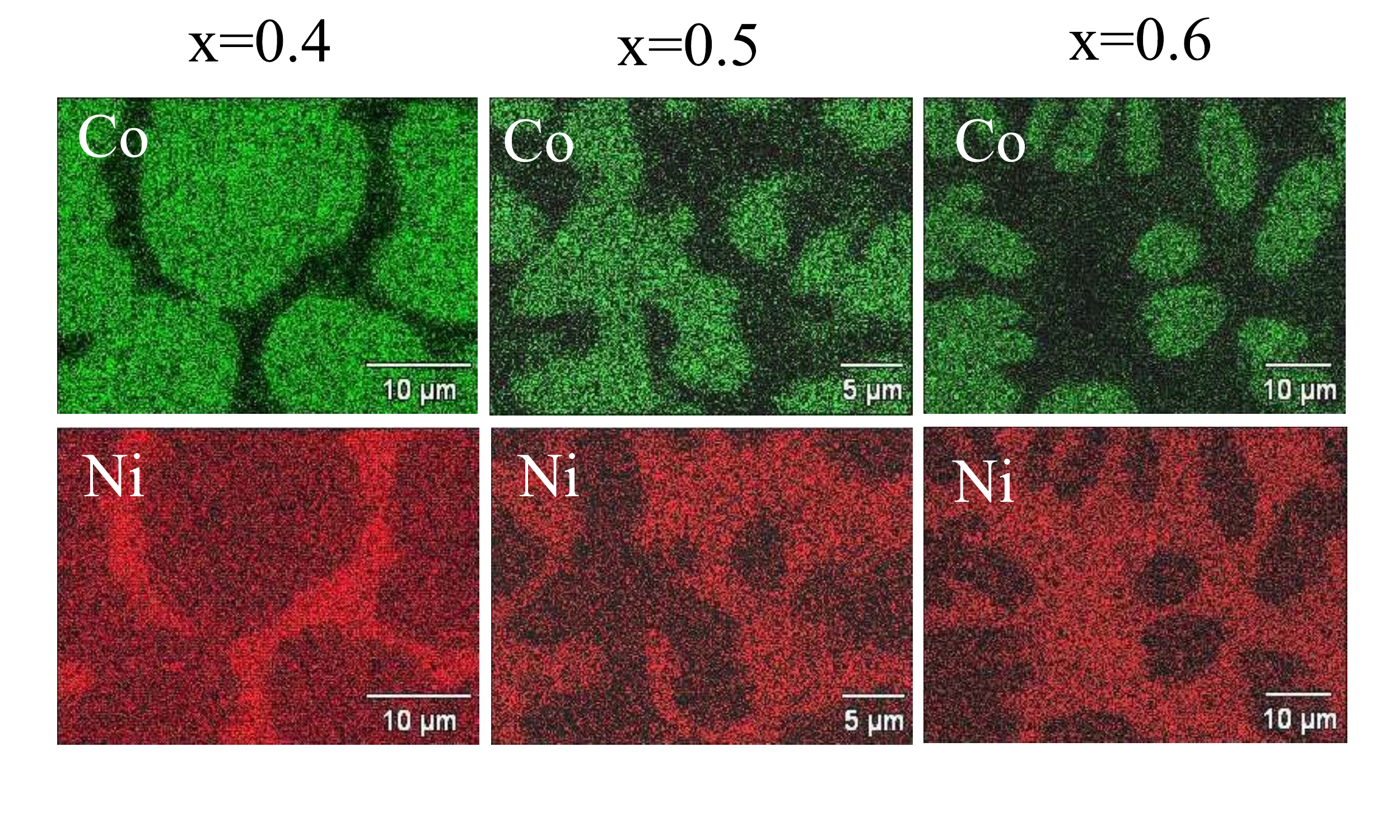}\\
  \caption{The images correspond to Co (green) and Ni (red) EDX maps for different values of x in Co$_{(1-x)}$Ni$_x$Si after annealing of the samples at 900$^oC$ during a week.} \label{microscopie}
\end{figure}

Between these two limits (for $0.35<x<0.74$), the two solid solutions at the limit of solubility coexist. The Fig.~\ref{microscopie} shows EDX Co and Ni maps in samples with nominal compositions $x=0.4$, $0.5$ and $0.6$.

The chemical composition and grains can be easily identified revealing sharp grain boundaries. The fact that cobalt rich grains are closed volumes comes from the fact that B20 cobalt rich structure has a higher melting point as we will see later. As a result, these grains nucleate and start growing first during cooling. B31 is growing at lower temperature in the free space between B20 grains.

\begin{figure}[h!]
   \center\includegraphics[width=12cm]{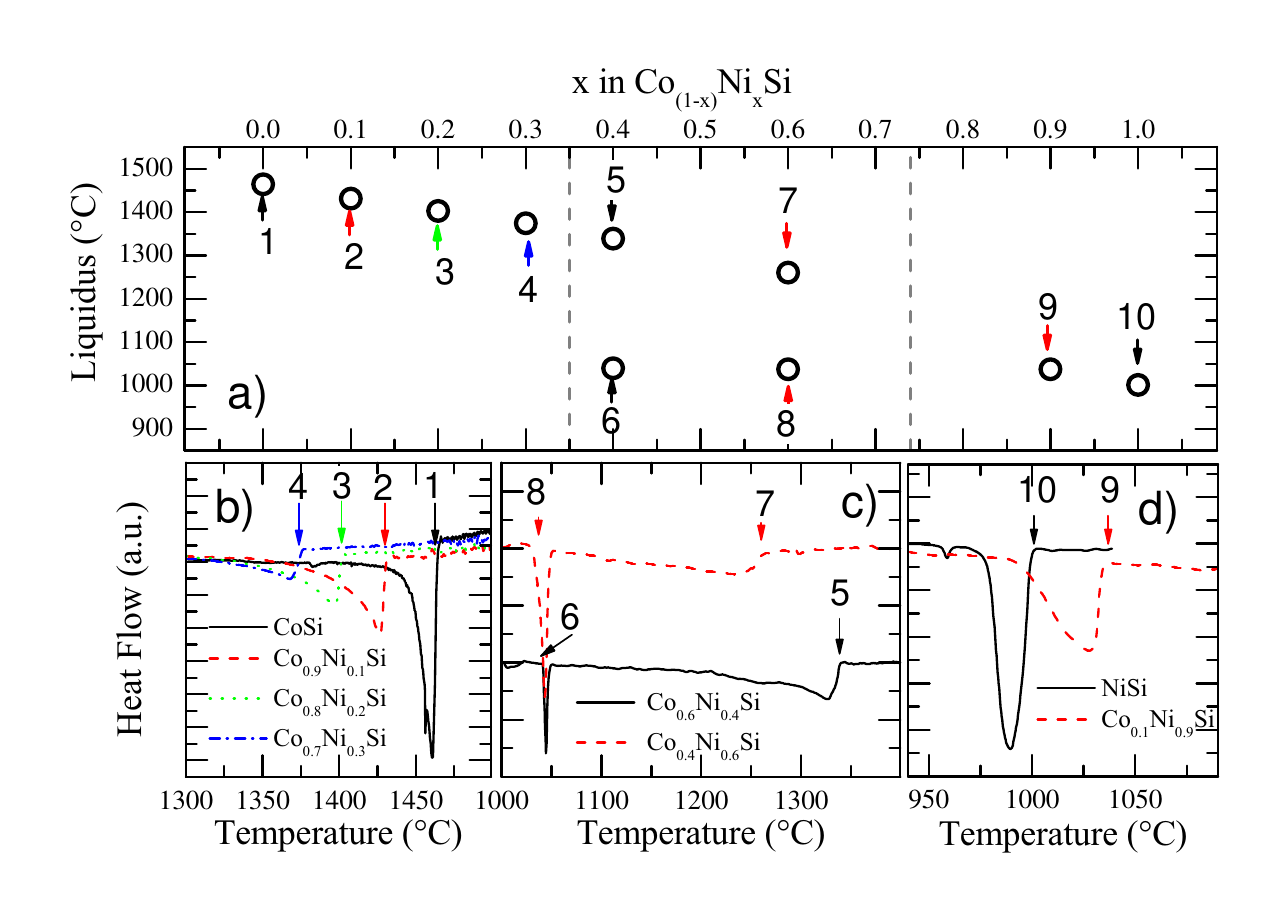}\\
  \caption{a) Evolution of the liquidus temperatures with x value in Co$_{(1-x)}$Ni$_x$Si. Limits of solubility are displayed as dashed gray lines. b-d) thermal flow curves for respectively B20, mixed and B31 solid solutions.} \label{DTA}
\end{figure}

This difference in melting temperatures was observed by thermal analysis for pure solid solution (Fig.~\ref{DTA}b and d) and for multiphase samples (Fig.~\ref{DTA}c). The evolution of the liquidus temperature with x is plotted in Fig.~\ref{DTA}a.

The melting temperature of NiSi $T=978^oC$ agrees well with other recent DTA experiments ($T=979^oC$ \cite{Richter2004}, $T=982^oC$ \cite{Du1999}). Whereas no recent data are available for CoSi, the melting temperature $T=1446^oC$ was found in the range of two old experimental reports ($T=1420^oC$ \cite{Vogel1934} and $T=1460^oC$ \cite{Haschimoto1937}).

These thermodynamic investigations first revealed that the liquidus temperature of these solid solutions decreases with $x$ in the two region presenting a single phase. In the multiphase region ($0.35<x<0.74$), two samples have been measured (Co$_{0.6}$Ni$_{0.4}$Si and Co$_{0.4}$Ni$_{0.6}$Si) (Fig.~\ref{DTA}c) pointing out a monovariant line at a temperature $T_i\sim1040^oC$ shown by points 6 and 8 in Fig.~\ref{DTA}a. The second endothermic peak at higher temperature correspond to the liquidus.

The magnetic susceptibility of some samples was measured from room temperature down to 4 K using a SQUID magnetometer showing diamagnetic behaviors over the whole temperature and composition range.

\section{Conclusions}

We have reported the study of the stability range of the two phases (B20 and B31) forming in Co$_{(1-x)}$Ni$_x$Si. The whole range of nominal composition $0<x<1$ was investigated. Ab-initio calculation were used to optimize the structural parameters and predict the structural transition from total energy computation. Systematic structural and thermodynamic studies of Co$_{(1-x)}$Ni$_x$Si ($0<x<1$) revealed the existence of 3 different sample morphologies corresponding to 3 distinct regions of the CoSi-NiSi quasi-binary phase diagram:
\begin{itemize}
  \item{A single phased B20 solid solution with a limit of solubility $x_1=0.35$}
  \item{A single phased B31 solid solution with a limit of solubility $x_2=0.74$}
  \item{and, between these two limit composition, a mixture with the two solid solutions at there limits of solubility.}
\end{itemize}
In the mixed phase range, a monovariant line at a temperature of 1040 $^oC$ was found.

\section*{Acknowledgments}

This work is supported by the Swiss National Science Foundation through grant 200020-109588 and the National Center of Competence in Research (NCCR) "Materials with Novel Electronic Properties-MaNEP".

\end{document}